# Reexamining the Chain Fountain

Hiroshi Yokoyama[1]

Advanced Materials and Liquid Crystal Institute,

Department of Physics, and Chemical Physics Interdisciplinary Program

Kent State University

1425 Lefton Esplanade, Kent, OH, USA

**ABSTRACT**

We show via proof-by-construction that the chain fountain, also known as the Mould effect, is a generic consequence of energy conserving dynamics of linear chains, similar to the extra acceleration observed for hanging or falling chains. Extracting a chain from its unentangled slack sitting on a table is much less energy dissipative than the claim of the classical scenario, postulated by Biggins and Warner[Proc. R. Soc. A **470**, 20130689(2014)] as the starting point of their explanation of the chain fountain. As a result, their alleged upward kicks on the chain from the table as the chain is uplifted can at best be one of the possible factors in driving the fountain, if not totally irrelevant. We construct an efficient chain fountain with no upward kicks in Biggins and Warner's sense, rather relying on the hanging chain dynamics. Simply put, the centrifugal force at the top of the fountain is responsible for maintaining the fountain. We argue that lateral motion of the chain in the slack plays a decisive role in the Mould effect.

## 1. Introduction

The dynamics of chains or inextensible strings has long been considered a typical example of variable mass systems in classical mechanics[1]. A vertically stretched chain freely falling on the floor, for example, is treated in such a way that the lowest portion of the falling chain abruptly stops on touching the floor without affecting the motion of the forthcoming or preceded portion of the chain. To make up for the lost momentum of the chain per unit time $\rho v \times v$ with $\rho$ being the density of the chain per unit length and $v$ the velocity, the floor has to exert an extra upward force $F_e = \rho v^2$ in addition to the weight of the chain already landed on the floor $F_g = Mg$. With the constant gravitational acceleration yielding $v = gt$ and $M = \frac{1}{2}\rho g t^2$, the total upward force from the floor is given by $F_{total} = F_g + F_e = 3Mg$. The kinetic energy of the chain is instantaneously dissipated at the impact of the chain on the floor. The same line of arguments is applied to the reverse process of extracting a chain at a velocity $v$ from a slack sitting still on a table. The momentum conservation requires that the force needed to abruptly set a chain in motion is $F = \rho v^2$; Since the kinetic energy the chain acquires is only half of the work done by the force, the rest half of the work must be dissipated at the point of extraction.

For the last few decades, however, mounting evidence has been accumulated both theoretically and experimentally that the real chains are much less dissipative and can be better approximated as an energy conserving system[2-8]. Consider a falling chain as illustrated in Fig.1, which consists in a chain hanging in the gravity with one end is fixed, and the other end is released. The classical argument assumes that the released end of the chain undergoes a free fall at the velocity $v = gt$ at time $t$ after the release, while the chain near the bottom continuously transferred from the moving to the static side at the rate of $v/2$. Through the transfer process, the kinetic energy is fully and instantaneously dissipated. Real observations as shown in Fig.1, however, show that the released end of the chain falls faster than the free fall. This so happens because the kinetic energy is not completely dissipated, but is at least partly recycled to accelerate the moving part of the chain. This is similar to the behavior of a whip. In the example shown in Fig.1, nearly complete conservation of energy is achieved. That the classical treatment is unrealistic may be intuitively obvious if we imagine the behavior of chains just after the fall. If the kinetic energy is fully and continuously dissipated along the fall as assumed, the chain should remain vertically still at the end of the fall. On the contrary, what we see is that the chain bounces vigorously on its own, clearly indicating a substantial amount of kinetic energy remaining undissipated.

---

[1] Author for Correspondence: hyokoyam@kent.edu

The chain fountain is a fascinating recent example of extraordinary behaviors that a simple chain can exhibit[9]. A long chain is put in a cup and the end of the chain is pulled down over the rim of the cup (see Fig.2). As the chain falls toward the floor by gravity, the moving chain is lifted up spontaneously over the rim, forming an arch not in contact with the cup. Since Mould presented a video entitled "Self siphoning beads" in 2013, this gravity defying phenomenon has attracted considerable attention both from the scientific community [10] and from general public[11,12].

Biggins and Warner argued, based on the classical treatment of the chain as being dissipative, that an upward impulsive force from the cup/table, associated with the collision of the chain with support at the point of uplifting, is necessary to make the fountain possible by complementing an otherwise insufficient upward momentum of the chain when extracted[10]. This explanation has been widely spread and is presently most trusted. Andrew et al. reported the dependence of fountain height on the drop distance up to 18m[13], and Pantalenone carried out a quantitative study of chain fountain along the Biggins-Warner scenario[14]. Martins, however, presented a few observations of chain behaviors that are apparently in conflict with the Biggins-Warner explanation[15]. More recently, Flekkøy, Moura and Måløy[16] also shed some light of skepticism mostly based on a simulation for different types of chains and surface conditions of the container.

In this paper, we demonstrate via proof-by-construction that the upward kick from the cup/table, even if it exists, is not always necessary for the formation of chain fountain. We show that the chain fountain can be more generally understood in a wider context as another *counter intuitive* behavior of chains resulting from their energy conserving characteristics in parallel with the falling chain phenomena.

## 2. Chain fountain without kicks: proof by construction

### 2.1 Simple analysis of hanging chains

As the first step, we consider the motion of a hanging chain that is approximated as two vertical linear chains connected by a semi-circular bottom of radius $R$ illustrated in Fig.3. As the hanging end of the chain is vertically raised ($v > 0$) or lowered ($v < 0$) at the velocity $v$, the motion of a point on the semi-circle $\boldsymbol{r}$ fixed on the chain follows

$$\boldsymbol{r} = \mp \left( R\cos\frac{v}{2R}t \quad \frac{v}{2}t + R\sin\frac{v}{2R}t \right), \tag{1}$$

where the minus and plus signs correspond to the rising and the lowering processes, respectively. The point is smoothly accelerated ($v > 0$) or decelerated ($v < 0$) while it is traversing the semi-circle (($0 \leq t \leq 2R\pi/v$). As can be readily confirmed on an assumption of constant velocity, the force responsible for the acceleration and deceleration is consistently provided by the following extra tension on the chain in addition to the tension to balance the gravitational force:

$$T_e = \frac{1}{4}\rho v^2, \tag{2}$$

where $\rho$ stands for the density of the chain per unit length. Note that no energy is dissipative through this process, since the kinetic energy acquired or lost per unit time is $\frac{1}{2}\rho v^2 \times \frac{1}{2}v = Tv$, i.e. the work done by or to the tension. In the case of falling chain, this tension generates the extra acceleration of the falling part to exceed the free fall by gravity. The degree of dissipation that may exist in real chains, especially for a small $R$, can be phenomenologically incorporated as

$$T_e = \frac{1}{4}\rho v^2 + \Delta, \tag{3}$$

where $\Delta$ is a velocity-dependent parameter, which is positive where the chain is exiting from the semi-circle, and is negative where the chain is entering the semi-circle; the energy dissipation per unit time is then given by $v\Delta$. In view of the fact that $\Delta$ changes sign according to the sign of $v$, the lowest order term in the power expansion of $\Delta$ with respect to $v$ must be linear:

$$\Delta \approx \pm \gamma v. \tag{4}$$

Here $\gamma$ is a friction coefficient, which is a positive parameter, dependent on $R$. The plus and minus signs apply to the moving and the static side of the chain, respectively. The observation of a hanging chain shown in Fig.1 is consistent with the extra tension without dissipation as given above.

### 2.2 Minimal chain fountain

Using the hanging chain, we now devise a minimal fountain as follows. As illustrated in Fig. 4, one end of the chain is fixed and the other end is to be released over a sliding rod. When the velocity of the chain is below a certain threshold, the chain is in contact with the sliding rod. At higher velocities, the centrifugal force on the chain along the sliding rod is enough to detach the chain from the rod and sustain the weight of the vertical chains on its own; Calkin considered the motion of a free chain falling over a sliding rod as a related yet separate problem from the hanging chain[3]. Assuming the energy conservation and $R_1, R_2 \ll h_1$, the balance of force on the part of the chain moving upward after detachment reads

$$\rho(g + \dot{v})(h_1 + h_2) = T_0 - T_e = \frac{3}{4}\rho v^2. \tag{5}$$

Here, the last equality holds only at the time just after the chain is detached from the sliding rod where we can assume $T_0 = \rho v^2$ and $T_e = \rho v^2/4$. At later times, both $T_0$ and $T_e$ should vary as the top is further lifted, while satisfying the first equality of Eq.(5).

The condition for the chain fountain to occur is $h_2 > 0$, which yields

$$v > v_{th} = \sqrt{\frac{4(g + \dot{v})}{3}h_1}. \tag{6}$$

This indicates that a chain fountain can occur from this minimal arrangement, provided the chain is accelerated to a sufficiently high velocity.

We carried out a simple experiment to demonstrate the validity of this prediction. The ball chain used is comprised of 2.8mm diameter nonmagnetic metallic balls flexibly linked with thin metallic 0.4mm-diameter rods, forming a chain of 254 balls/m and the density 21.3g/m.  As the sliding rod, a 22mm diameter acetal rod was used for its excellent rigidity and low friction.  One end of the chain is fixed at 185cm above the floor and the sliding rod was held at the height of 128cm from the floor using a thin copper wire.  The total length of the chain is 185cm, so the chain just touches the floor when fully extended. The initial length of the hanging part is $h_1(0)$ =60cm. Instead of releasing the chain alone, a heavy stainless steel weight ($M_w$ =250g) was attached to the chain and dropped from the position of the sliding rod to the floor in order to better regulate the velocity regardless of the conditions of the chain.  Since the weight of the hanging part of the chain is indeed less than 5% of the attached weight, we can approximately assume that the velocity is given by the free fall equation $v = gt$ independent of $h_1$ and $h_2$.  Since $h_1 = h_1(0) - gt^2/4$ holds until the detachment, we obtain from Eq.(6) the time the chain fountain sets in as follows:

$$t_d = \sqrt{\frac{8}{5g} h_1(0)}, \tag{7}$$

which yields for the present condition $t_d$ =313msec. In terms of $L = gt^2/2$, the distance traversed by the dropped weight, Eq.(5) can be rewritten as

$$h_1(t_d) = \frac{3}{4} L. \tag{8}$$

Note that since the tension scales as $\rho v^2 = 2L\rho g$, the effect of tension on the falling velocity is also considered small compared to the gravitational force on the dropped weight as long as $L$ is not too long.

The motion of the chain was captured by using a high speed camera (fps1000, The Slow Motion Camera Company, Ltd.) at the rate of 1000 frames/s.  Figure 5 shows the typical time laps snapshots of the chain before and after the appearance of the fountain.  After 350msec after release, the chain on the sliding rod starts rising.  As Figs. 5(d), (e) and (f) show, the fountain further rises while the hanging part shrinks and goes even higher than the sliding rod.  In this arrangement, there is a geometrical constraint that keeps $h_1 + h_2 + L/2$ always constant.  Therefore, the height of the fountain $h_1 + h_2$ decreases as the weight falls and disappears when the weight reaches the floor.   The onset time of the chain fountain agrees with the above calculation.  The slight delay may be ascribed to the influence of dissipation and/or the ignored lengths in the semi-circles.  For more details, watch the high speed video in supplement.

Measuring the dropped distance of the weight as a function of time, we confirmed the free fall assumption was valid in the present experiment (Fig.6).  Figure 7 shows the height $h_2$ as a function of time as measured from the time lapse images, present clearer evidence about when the detachment of the chain took place.  The results indicate that the behavior of the minimal chain fountain system obeys rather closely the predictions made without dissipation.  Since the dissipation is linear to the velocity as in the case of friction, Eq.(4), the tension along the chain tends to be more dominated by the inertial contribution proportional to $\rho v^2$ as higher velocities.  Even for a short chain used here wherein the highest velocity is limited to 6m/s, we can conclude that the dissipation is causing a minor effect.

Finally, it is worth pointing out that the beam to which the end of the chain is fixed starts exerting an extra upward force $\frac{1}{4}\rho v^2$ (or more precisely $\frac{1}{4}\rho v^2 - \Delta$ including dissipation) as soon as the chain is set in motion. This force does not do any work on the chain, but is needed to provide a part of upward momentum of the chain.  Its role in chain fountain is identical to that of the upward kicks suggested by Biggins and Warner[10], but its origin is completely distinct.

### 2.3 Chain fountain without upward kicks

The minimal chain fountain can exists for only a fraction of a second in an ordinary laboratory environment; so it is not possible for ordinary humans to recognize it with naked eyes. We devised a novel chain fountain system by cascading multiple minimal chain fountains in a circular configuration (Fig. 8) in such a way that the chain fountain is sustained for a sufficiently long period of time, yet still lacks in any mechanisms to support the kick-driven fountain scenario. We used a 0.25mm-thick stainless steel sheet to form a crown shaped support as shown in Fig. 8: The height of the crow is 8cm and the diameter is 19cm. The 9.5m-long chain was placed around the rim of the crown with alternate short hanging parts inside and outside the crown. The crown is in contact with the chain only at the thin edge of the metal sheet, on which the linkage between the balls settle. Except this point, there are no other contacts with the chain and the support. To aid smooth start of the chain fountain without initial bumpy collisions of the ball with the crown edge, the beginning 7cm of the rim is covered with a 7mm-diameter stainless tube, over which the first three or four hangings reside. The part of the chain connected to this launching section (the chain slack seen inside the crown) is released inside the crown toward the floor. The crown was placed at 123cm above the floor on a beam having a center opening to allow the chain to fall.

Figure 9 shows the snapshot of the chain fountain from the crown. Thanks to the short launch pad, the fountain shows up almost immediately after the release of the chain. It grows as the hanging segments of the chain are consumed. The picture in Fig. 9 was taken halfway through the rim by a high speed camera at 1000frames/s. Following the red markings on the chain that were placed at every 50cm, we can directly measure the falling velocity of the chain. Shown in Fig. 10 is the temporal variation of the velocity and the fountain height; for comparison, the results for Mould's chain fountain shown in Fig.2 are included.

The velocity reaches a steady value of $v$ =4.2m/s in about 500ms after the chain starts falling, which is about the same falling speed as in Mould's chain fountain. The final velocity of a free fall from 123cm-high point is $v_{free}$ =4.9m/s, which gives a ratio $v/v_{free}$ =0.85 in consistent with the previous results[10,13,14]. On the other hand, there is a significant difference in the height of chain fountain between Mould's fountain and that from the crown. The highest fountain height attained in Mould's chain fountain is 16cm, but the chain fountain from the crown reaches 25cm, indicating the superior efficiency of the crown arrangement. In the absence of acceleration, the height of the fountain $h_2$ is approximately related to the difference of tension between the top and bottom of the fountain by $\rho g h_2 = T_o - T_e$. In the completely dissipationless case with $T_e = \rho v^2/2$, it follows from $T_o = \rho v^2$ that $h_2 = v^2/2g$. Using the observed velocity, it predicts $h_2$ =90cm, which is way larger than the actual fountain heights.

The chain fountain from the crown consists of repetition of two stages: (1) Minimal fountain stage and (2) Abrupt mass increase stage. When the minimal fountain consuming a small hanging part comes to an end, the fountain moves to the next hanging portion on the other side of the crown. At this point, the support from the crown edge suddenly disappears and half of the mass of the next hanging chain, which is still immobile, is directly connected to the fountain. The added mass must be abruptly accelerated to $v$; this process is exactly what is assumed in the classical picture of the chain dynamics, so it dissipates the energy $h_1 \rho v^2/2$ where $h_1$ is the depth of the hanging chain. Following this dissipative process begins the next minimal chain fountain accelerating gradually the $h_1$-long chain to $v$ without energy dissipation. On the average, therefore, the tension at the bottom of the fountain from the crown is given by

$$T_e = \frac{3}{4}\rho v^2. \tag{9}$$

We then find that the theoretical limit of fountain height from the present crown is 45cm. This is still too large compared to the observed 25cm; however, given the neglect of dissipation in the minimal fountain and at the transition from one hanging part to the next, and also the crude approximation of the fountain shape such as the straight chain as opposed to the wiggled form in actual fountains, the discrepancy seems satisfactorily small.

For Mould's chain fountain, there is no straightforward and realistic argument to fill the gap between the theoretical 90cm and 16cm observed in reality. As will be discussed in the next section, the lateral motion of the chain perpendicular to itself seems to be playing a key role.

## 2.4 Chain fountain as a reverse process of falling chain slack

The above implementation does not definitively answer what is actually driving the chain fountain from the slack sitting in a cup or on a table. Since existing chain fountains satisfy all the conservation laws regardless of the underlying mechanisms, any observation of macroscopic variables such as the mass, force and tension cannot address the cause of chain fountain, unless finer details of the chain motion inside the slack are elucidated at the ball-by-ball level; this is particularly true for examining the kick-driven scenario[10]. The upward force from the support in excess of the weight of the remaining chain always exists for the sake of total momentum balance whatever the real mechanism of uplifting is, so the presence of this force cannot be evidence for any model.

Here, in order to shed light from a different angle, we tried a reverse process of the chain fountain by letting the slack fall freely with the top end of the chain fixed as illustrated in Fig.11. Although it was a little tricky, it turned out to be reproducibly possible to quickly pull down the support plate (depicted by a broken line in the figure) by hand to initiate the free fall. To compare the result with the real free fall in gravity, the same aluminum weight as used before was simultaneously released.

The main question to answer is how much potential energy be dissipated through this fall. Since the chain slack has no container or support, there is no way for the alleged upward kicks to help reduce the energy dissipation. According to the classical "fully dissipative" chain picture, there must be an extra tension $\rho v^2$ on the immobile hanging chain emanating from the slack. Since the end of the chain is fixed, this tension does not do any work on the chain, but is only necessary to keep the momentum conservation when the falling chain is continuously stalled. The kinetic energy per unit length of chain $\rho v^2/2$ is completely dissipated when the chain is transferred from the slack to the hanging chain. Hence, the slack falls at the same speed as the true free fall in gravity and all the kinetic energy is lost when the slack is fully extended. In other words, at the end of the free fall of the slack, we have to have a still vertical chain.

Figure 12 shows the time laps pictures of the falling chain slack and the aluminum weight. Within the experimental error, the slack falls at the same velocity as the aluminum weight. A special care was always taken to avoid tangling the chain in the slack, which results in a noticeably delayed fall. For numerous trials, we have never encountered a situation that the slack falls significantly faster than the weight unlike the case of released hanging chain shown in Fig.1. A striking difference from the classical scenario, however, is that at the completion of free fall, the chain still retains a substantial portion of the acquired kinetic energy as manifested by the strong rebound and vigorous vibration of the chain afterwards.

Although it is difficult to accurately quantify the remaining energy, we can make a rough estimate by analyzing the velocity and position of the rebounding chain from the snapshots between the completed fall (Fig. 12(c)) and its top rebound position (Fig. 12(e)). For this purpose, we used an image analysis software ImageJ distributed for free from National Institute of Health (https://imagej.nih.gov/ij/). The total potential energy difference of the chain before and after the free fall is $E_p = \rho g H^2/2$ with $H$ =180cm. The estimated remaining energy, the sum of kinetic and potential energies, ranges between 5%~10% of $E_p$. The deviation from the classical fully dissipative model of chains requires that the tension on the chain at the slack should be reduced by $\alpha \rho v^2/2$ from the classical tension $\rho v^2$, where $\alpha = 1$ corresponds to the case of complete conservation of energy. In the present free fall situation, the work this reduced tension does on the slack is:

$$W = \int_0^{t_c} \alpha \frac{1}{2} \rho v^2 \, v dt = \alpha E_p, \tag{10}$$

where use is made of $v = gt$ and $H = gt_c^2/2$. Therefore, the above estimate of the rebound energy yields $\alpha = 0.05 \sim 0.1$. The height of the chain fountain is related to $\alpha$ by $\alpha = h_f/H_f$ where $H_f$ is the height of the fountain from the floor. For a variety of Mould's chain fountains including that shown in Fig.2, it has been consistently observed that $h_f/H_f \sim 0.14$[10,13,14]. Given the crudeness of the estimate of $\alpha$ from the dropped slack, we find the agreement between the values attained via different routes seems reasonable.

With the reduced dissipation, how can we reconcile the finite downward force from the vertical chain and the observed absence of extra acceleration in the falling slack, in contrast to the simple falling chain presented in Fig.1? One of the conspicuous differences between the falling chain slack and the simple falling chain is that the hanging chain left behind remains almost completely straight and vertical in the case of simple falling chain (see Fig.1 and its full video); whereas the hanging chain associated with the falling slack is very much agitated even before the end of chain reaches the floor. Let us now calculate the expected time difference of arrival, assuming that the motion is restricted in the vertical direction. The equation of motion for the slack is written as

$$\rho L_s \frac{dv}{dt} = \rho L_s g + \alpha \frac{1}{2} \rho v^2, \qquad (11)$$

where $L_s$ is the length of chain left in the falling slack. Noting that $dL_s/dt = -v$, we can integrate the above equation as

$$v^2 = \frac{2gL_s}{1+\alpha}\left\{\left(\frac{L_{s0}}{L_s}\right)^{1+\alpha} - 1\right\}, \qquad (12)$$

where $L_{s0}$ stands for the initial length of chain in the slack at the time of release. The time necessary for the chain slack to reach the floor is then given by

$$t_f = \sqrt{\frac{2L_{so}}{g}} \int_0^1 \frac{1}{2}\sqrt{\frac{(1+\alpha)x^\alpha}{1-x^{1+\alpha}}}\,dx. \qquad (13)$$

The integral gives the ratio of the required time and the time of free fall $t_{free} = \sqrt{2L_{x0}/g}$. When $\alpha = 1$ corresponding to the complete energy conservation, $t_f/t_{free} \approx 0.85$. For smaller values of $\alpha$ below 0.2, numerical integration confirms an approximate relation $t_f/t_{free} \approx 0.2\alpha$. For the present experimental conditions, $t_{free} = 600$ms and we find the predicted time gap of arrivals for $\alpha = 0.05 \sim 0.1$ to be $t_{free} - t_f \approx 6$ms $\sim 12$ms. Since the velocity at arrival is 5.9m/s, the expected distance between the chain front ant the weight is 3cm $\sim$ 6cm. In view of 1ms time resolution of the imaging system we used, this level of time and distance gap should be reliably detectable. We are therefore compelled to consider where the saved kinetic energy goes.

As illustrated in Fig.13, the lateral extension and the horizontal arrangement of chain inside the slack excites a motion of the chain in the direction perpendicular to itself. The kinetic energy is distributed not only in the vertical direction but also in the lateral degrees of freedom that do not participate in the motion in the simple falling case. Consequently, it is naturally anticipated that the extra acceleration in the downward direction should be less effective in the falling chain slack compared to the simple falling chain.

The falling chain slack experiment has demonstrated that the extraction of chain from the slack is inherently less dissipative than the classical postulate that unduly urged Biggins and Warner to invoke the aid of upward kicks. The chain fountain is now simply understood as the reverse process of the falling slack, in which the chain slack is held on a table with the end of the chain being pulled upward. Unlike the previous consideration of chain fountain phenomena[10,13,14], we emphasize the critical importance of lateral motion of the chain when the chain departs from the slack[15,16]. If unrestricted, the lateral motion would eat up the saved energy of extraction, making it impossible or much less efficient to drive the chain fountain. This view is consistent with the recent simulation results [16], the observation of inefficient chain fountain from a unidirectionally coiled chain slack[15], and the formation of a broad arch when the chain is driven laterally from a single layer of folded chain on a flat table [17]. It also seems compatible with the finding that the fountain height grows increasing slowly at larger drop distances [13]; indeed, a higher velocity of chain would make it more difficult to restrain the lateral motions in the slack, thereby providing more kinetic energy in the motion perpendicular to the chain.

### 2.5 Chain fountain from tailored slack

As a last piece of experimental evidence for energy conservation in chain fountains, we present in Fig.14 a chain fountain from a regularly arranged pile of alternating layers of x and y oriented chain on a 100mm-square corrugated plastic plate. The chain was folded periodically into a linear array of eight lines with 180degree turn at every 8cm alternating layer-by-layer in x and y directions. As Fig.14(a) shows, there occurs a 10cm or so high chain fountain just by falling the chain over the edge of the plate. The uplifting point migrated over the slack, but only within a restricted central region in x and y directions according to the direction of the chain. This behavior is combined with a precursory acceleration of chain inside the layer up to at least 1m/s prior to liftoff (Figs.13(b)-(d)). The preparatory movement of balls before takeoff was reported previously in Ref.[16] on a slack in a cup as well. The dissipation in the classical model comes from the discontinuous acceleration at the point of extraction. Virga developed a formal theoretical expression for the dissipation in terms of the degree of velocity discontinuity along the chain across the liftoff point, taking into account the balance of lateral as well as vertical forces[18]. Roughly speaking, the rate of maximum energy dissipation at an abrupt chain extraction is written in terms of the velocity discontinuity as

$$E_d = \frac{1}{2}\rho(v - v_p)^2 \times v, \tag{14}$$

where $v_p$ is the velocity of the chain right before extraction. Using $v$ =4.1m/s and $v_p$ =1m/s found in Fig.14, Eq.(14) predicts 40% reduction of dissipation from the fully dissipative case ($v_p$ =0). This is more than enough to account for the reduced tension at the liftoff point to support the height of fountain observed.

It may be in order to discuss how the lateral motion of chain in the slack impacts the understanding of chain fountain. As clearly seen in Figs. 2, 9, and 14, and their slow motion videos, the wiggling near the liftoff point is a common behavior of the chain fountain, which results from the well-known generic instability of a chain or string moving at high speeds in the negative dynamic tension regime ($T - \rho v^2 < 0$)[19]. Although one may be tempted to use an inverted catenary as the stationary form of chain fountains [10,15], the rising section of the chain fountain can never be stationary, but is vigorously migrating over the slack (see the video for Fig.14). Nevertheless, the falling side of the fountain may be reasonably approximated by an inverted catenary curve. Referring to Fig.15, let us consider the balance of force of a fictitious box embracing the entire chain above the table surface. Assuming the chain is falling in the page plane, the force balance in the vertical and the lateral directions are respectively given by

$$Mg = F_\perp - T_e\sin\theta + \rho v^2\sin\theta, \tag{15}$$

$$0 = F_\parallel + T_e\cos\theta - \rho v^2\cos\theta, \tag{16}$$

where $M$ is the mass of the chain inside the box, and $\rho v^2\sin\theta$ and $-\rho v^2\cos\theta$ are the loss of momentum per unit time due to the falling chain. Equation(15) indicates that the chain fountain is partly upheld by the reaction from the ejected falling chain like a rocket. Using the inverted catenary formula

$$T_e = \rho v^2 - \frac{\rho s_f g}{\sin\theta}, \tag{17}$$

with $s_f$ being the length of the falling chain inside the box, we obtain from Eqs.(15), (16) and (17) the forces from the table to the chain as

$$F_\perp = Mg + (T_e - \rho v^2)\sin\theta = \rho s_t g, \tag{18}$$

$$F_\parallel = -(T_e - \rho v^2)\cos\theta = \rho s_f g\cot\theta = R_t \rho g. \tag{19}$$

Here $s_t$ stands for the total length of chain on the rising side, and $R_t$ is the radius of curvature at the top of the fountain. Equation (18) is a trivial expression of the weight balance, which is always valid regardless of the specific microscopic mechanism of liftoff. Although Eq. (19) is also a trivial statement of force balance, it appears more illuminating regarding the underlying physics of chain fountain. We see from Eq.(19) that the higher and the more laterally extended the chain fountain becomes, the required lateral force from the table becomes larger. It is in fact a common experience in chain fountain experiments that when the chain is placed on a smooth flat table, the chain slack is pushed to the opposite side of the fountain, further making the fountain more extended, resulting in further acceleration of the backward push of the chain slack. Similar behaviors were seen in Ref.[17], which reported a broad arch from an extended single layer of chain on a table. To prevent this instability, it is found effective to put the chain slack in a cup or on a corrugated surface as the experiment shown in Fig. 14. Conversely, provided other conditions remain the same, Eq. (19) also indicates that decrease of $F_\parallel$ makes $\theta \to \pi/2$ and $R_t \to 0$, so the fountain becomes more upright and sharp.

Finally, we show via proof-by-construction again that dissipationless extraction of chain is possible, while allowing for the formation of chain fountain, i.e. $T_e < \rho v^2$. As illustrated in Fig. 16, we make use of a folded chain in a mode similar to the hanging chain; this is essentially identical to the crown fountain with the hanging chain laid flat on a table. The extraction occurs in two dimension in the x-z plane. The velocity of the extracted chain is $v$ in z-direction and $-v/2$ in x-direction, so the total kinetic energy per unit length is $3\rho v^2/4$. In the absence of dissipation, the tension at the liftoff is $T_e = 3\rho v^2/4$. While the extracted chain moves to the negative x-direction at the velocity of $-v/2$, the chain on the x-y plane moves oppositely at the velocity $v/2$. The chain is continuously accelerated from the immobile chain through a semi-circular trajectory moving in the positive x-direction at $v/4$. Since the acceleration is continuous, no dissipation associated with the abrupt acceleration in the classical model is involved. In this specific construction, the tension at extraction is exactly the same as that for the crown chain fountain (Eq.(9)). It is also worth noting that the area of migration of the liftoff point in this scheme is 1/2 in the linear dimension of the slack; this is similar to the situation shown in Fig. 14. One caveat is that the semi-circle trajectory is not necessarily a self-consistent stable trajectory for the given tension unlike the hanging chain. As a result, the folding and uplifting portions of the chain may be subject to deformation. This is also consistent with the behavior of chain fountain shown in Fig.14 (see also Ref.[16]). The wiggling instability is coupled with or driven by the lateral motion of the rising chain. In Fig.14, the periodic migration of the liftoff point is similar to shaking a whip by hand, and is exciting a wiggled structure in the rising chain. Once a wiggled or looped shape is formed, it remains remarkably stable owing to the conservation of angular momentum, and it acts like a fixed channel along which the chain is restricted to travel. The complete understanding of chain fountain entails detailed analysis of the precursory acceleration, migration of liftoff point, and generation and decay of wiggled shapes.

## 3. Conclusion

We have demonstrated via proof-by-construction that chain fountains can be formed without impulsive upward kicks from the container, and are more generally understood in the context of energy conservation in chain dynamics. Although the chain is rigid along its length, the motion of chain is not restricted in one dimension along the length, but a motion in the directions perpendicular to the chain is also allowed. The degrees of freedom in the orthogonal dimensions allow for gradual acceleration and deceleration of the chain, and help avoid extra energy consumption. The classical treatment of chain dynamics, claiming fully dissipative behaviors, rests on the one dimensional view of chains combined with the neglect of correlation between neighbors along the chain. In the present study, the hanging chain, which has a proven energy conserving characteristics, has been utilized as an element of the devised models of chain fountain. The crown chain fountain provided concrete evidence for a marginal effect of upward kicks even if they exist. Although the dynamics of chains in slack is more complex than hanging chains, it appears plausible to speculate that chains should find a route of motion in 3D space, causing the least dissipation of energy among the available options. In particular, we pointed out the critical role of the lateral motion of chain in chain fountains drawing on the drop experiment of chain slack. We believe that a certain structural mechanisms such as surface roughness to nondissipatively hinder lateral motion of the chain must be present in the chain slack. As already pointed out, it is not possible to discriminate the kick driven mechanism from other mechanisms only from macroscopic observations of chain fountains. In order to reach a complete understanding of Mould's chain fountain, detailed microscopic examination of the chain motion in the slack is required.

## Acknowledgement

The author is deeply indebted to Peter Palffy-Muhoray for drawing the author's attention to the chain fountain phenomenon. Inspiring discussions with him is also very much appreciated.

## Supplementary Information:

Fig.1
https://www.dropbox.com/s/omenbusnbp24bb0/Fig%201%20Falling%20chain.wmv?dl=0

Fig.2
https://www.dropbox.com/s/tevep4lpwggj95s/Fig%202%20Mould%20Chain%20Fountain.wmv?dl=0

Fig. 5
https://www.dropbox.com/s/mxxsxn3he7zxww9/Fig%205%20Minimal%20Chain%20Fountain.wmv?dl=0

Fig. 9
https://www.dropbox.com/s/7xiyj5jpnzooocd/Fig%209%20Crown%20Fountain.wmv?dl=0

Fig. 12
https://www.dropbox.com/s/516nvao514tmhrj/Fig%2012%20Falling%20Slack.wmv?dl=0
https://www.dropbox.com/s/din0bx9l70iyp7b/Fig%2012%20Falling%20Slack%2010ms-1sec.wmv?dl=0

Fig. 14
https://www.dropbox.com/s/xu7odyidv2f518n/Fig%2014%20Chain%20Fountain%20from%20XY%20Stack.wmv?dl=0

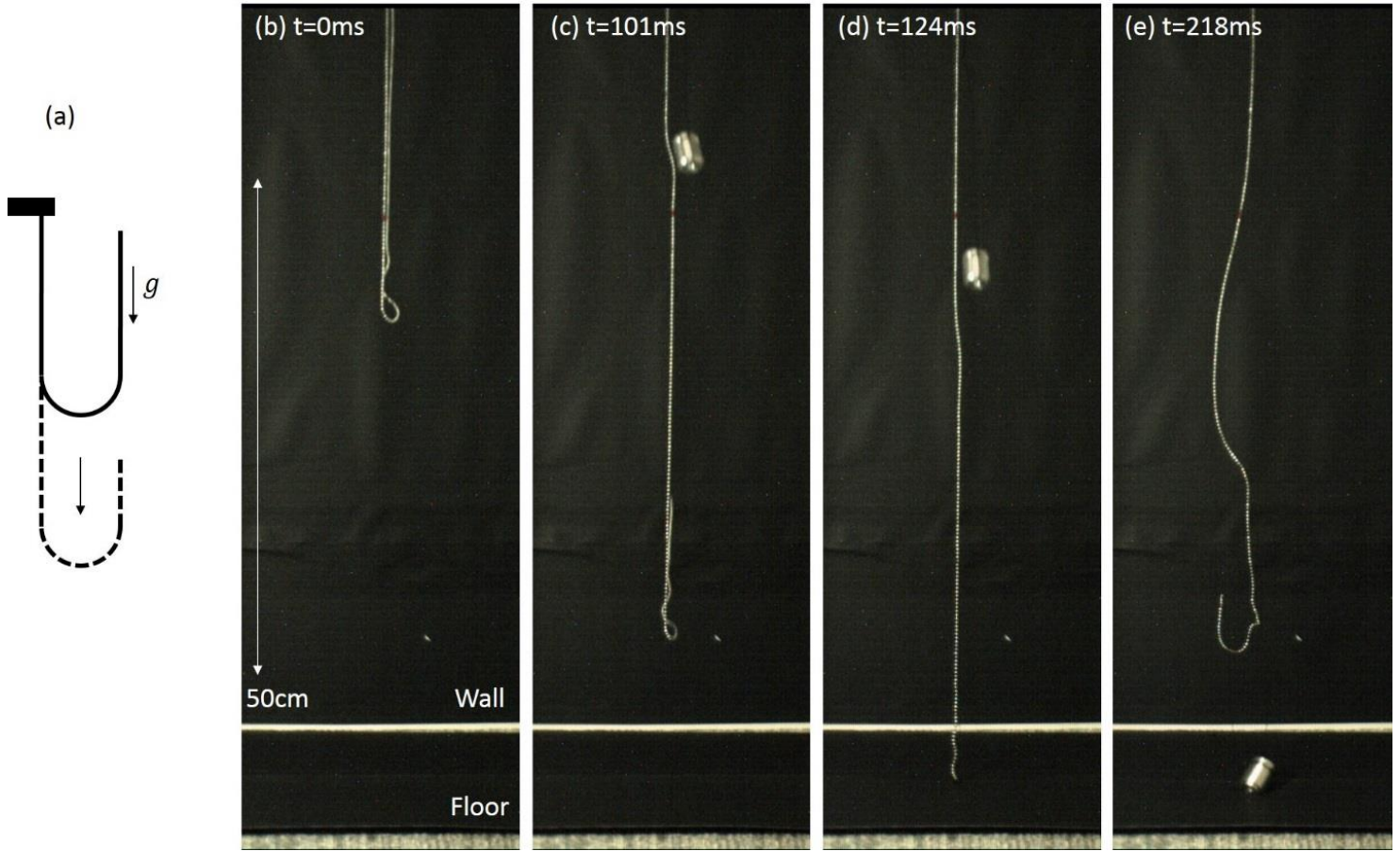

Figure 1. Extra acceleration of a released hanging chain. (a) One end of the chain is fixed and the other end is released in gravity to fall toward the floor. (b)-(e) Time laps snapshots of a falling chain. The free end of the chain was released together with a separated aluminum weight to compare the falling speed of the chain to the free fall in gravity. The total length of the chain was 185cm and one end was fixed at a rigid ceiling; the chain was folded in half with the other end held at the same height as the fixed point. The images were taken with a high speed camera (fps1000, The Slow Motion Camera Company, Ltd.) at the rate of 1000frames/s. The times indicated in the figures are relative time lapse from figure (b). In figure (c) both the chain and the aluminum weight are both still in the air, but it is clear that the chain is falling ahead of the aluminum weight. Figure (d) is exactly at the time the released end of the chain reached the floor. Figure (e) was taken when the aluminum weight reached the floor; the chain was already bounced back in the air. The time difference between the arrivals of the chain and the aluminum is 94ms, which is in agreement with the theoretical prediction of 91ms assuming the conservation of energy[2]. The complete slow motion video is available as supplemental information.

https://www.dropbox.com/s/omenbusnbp24bb0/Fig%201%20Falling%20chain.wmv?dl=0

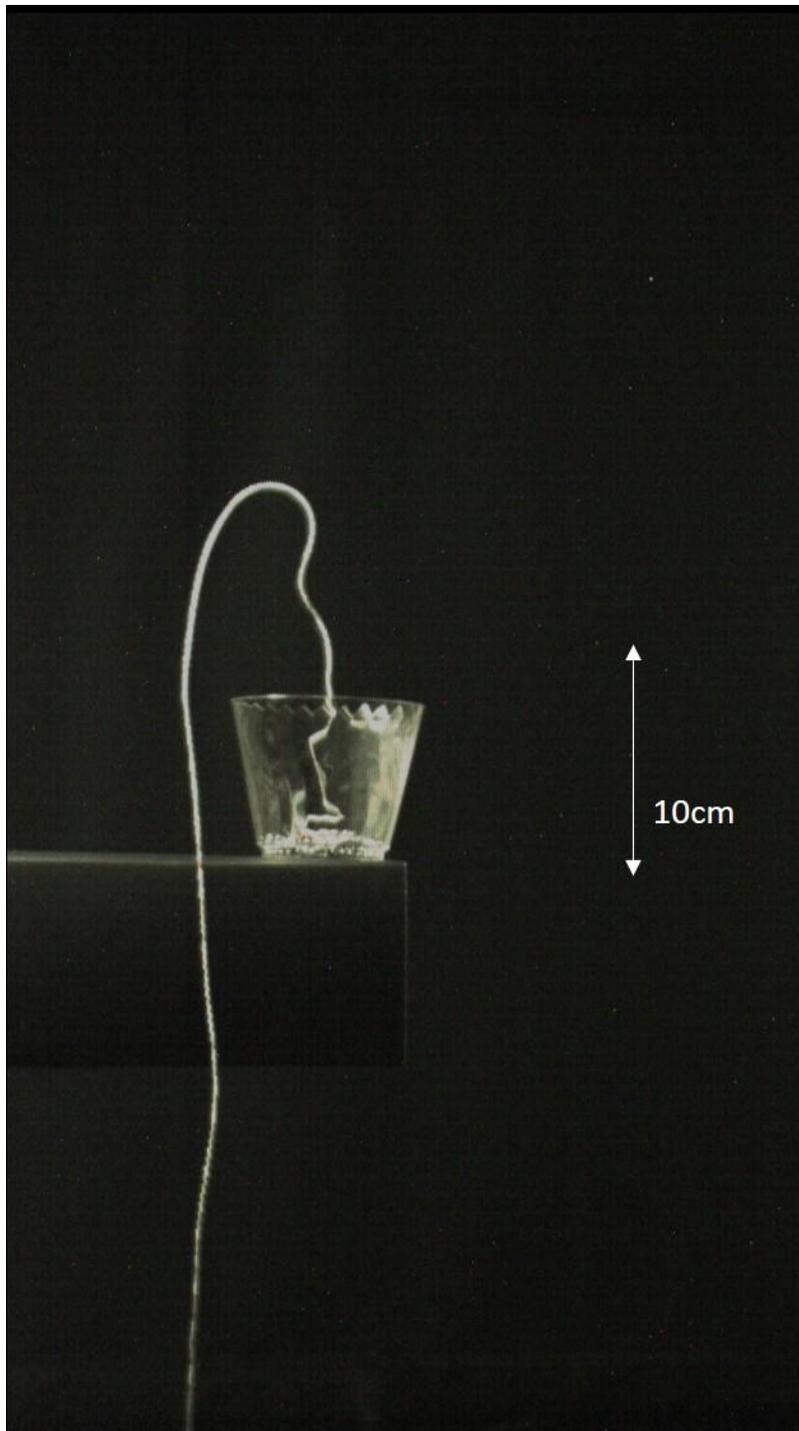

Figure 2. Mould's chain fountain from a slack in a plastic cup. The chain is falling to the floor 123cm below the cup. A snapshot captured by a high seed camera at the rate of 1000frames/s. The falling velocity is 4.2m/s, and the height of the fountain is 16cm. The complete slow motion video is available as supplemental information.

https://www.dropbox.com/s/tevep4lpwggj95s/Fig%202%20Mould%20Chain%20Fountain.wmv?dl=0

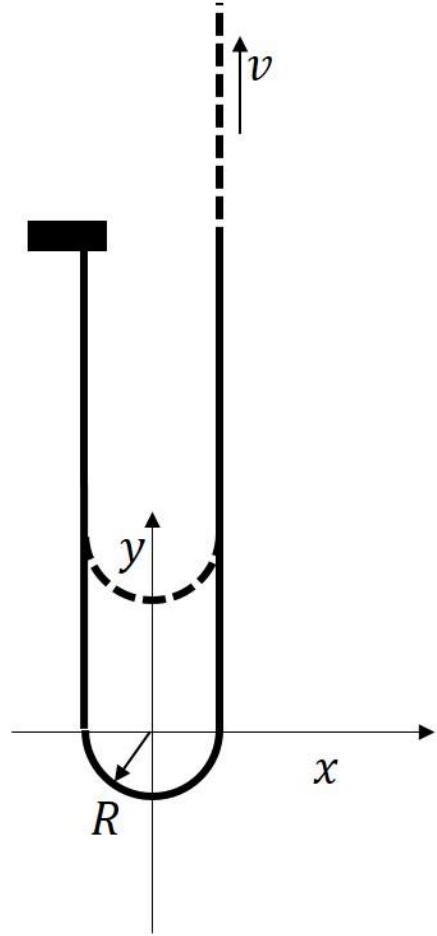

Figure 3. A hanging chain with one end is fixed on the beam and the other end is moving upward or downward at a constant velocity $v$. The chain is continuously accelerated from or decelerated to the immobile (left) part of the chain through the semi-circle part at the bottom. Motion away from the vertical axis, along with the chain is moving, allows for dissipationless extraction and settling of a chain, giving rise to non-classical behaviors of chains.

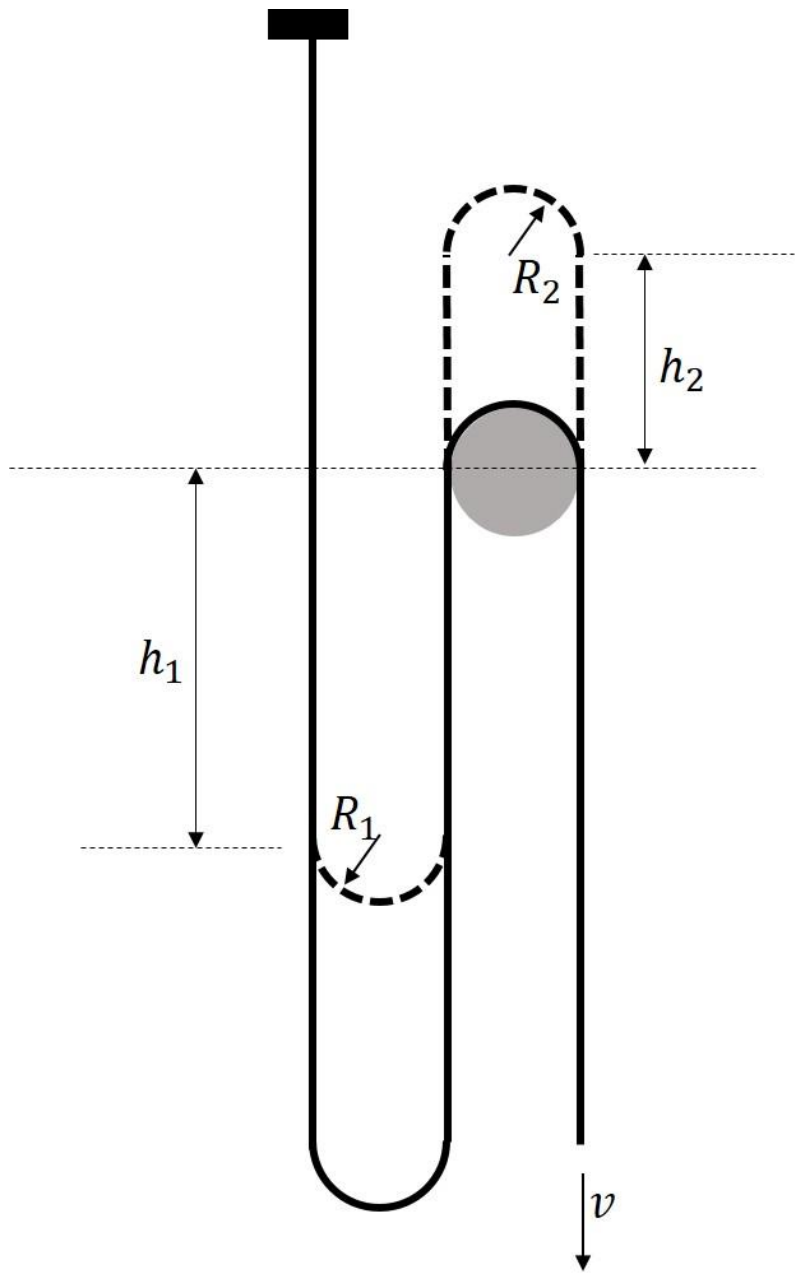

Figure 4. Configuration of a minimal chain fountain utilizing the hanging chain. The mobile end of the chain (right) is pulled down to the floor over a sliding rod, depicted in gray color. Beyond a threshold velocity, the centrifugal force on the curved portion over the sliding rod detaches itself from the rod and forms a chain fountain.

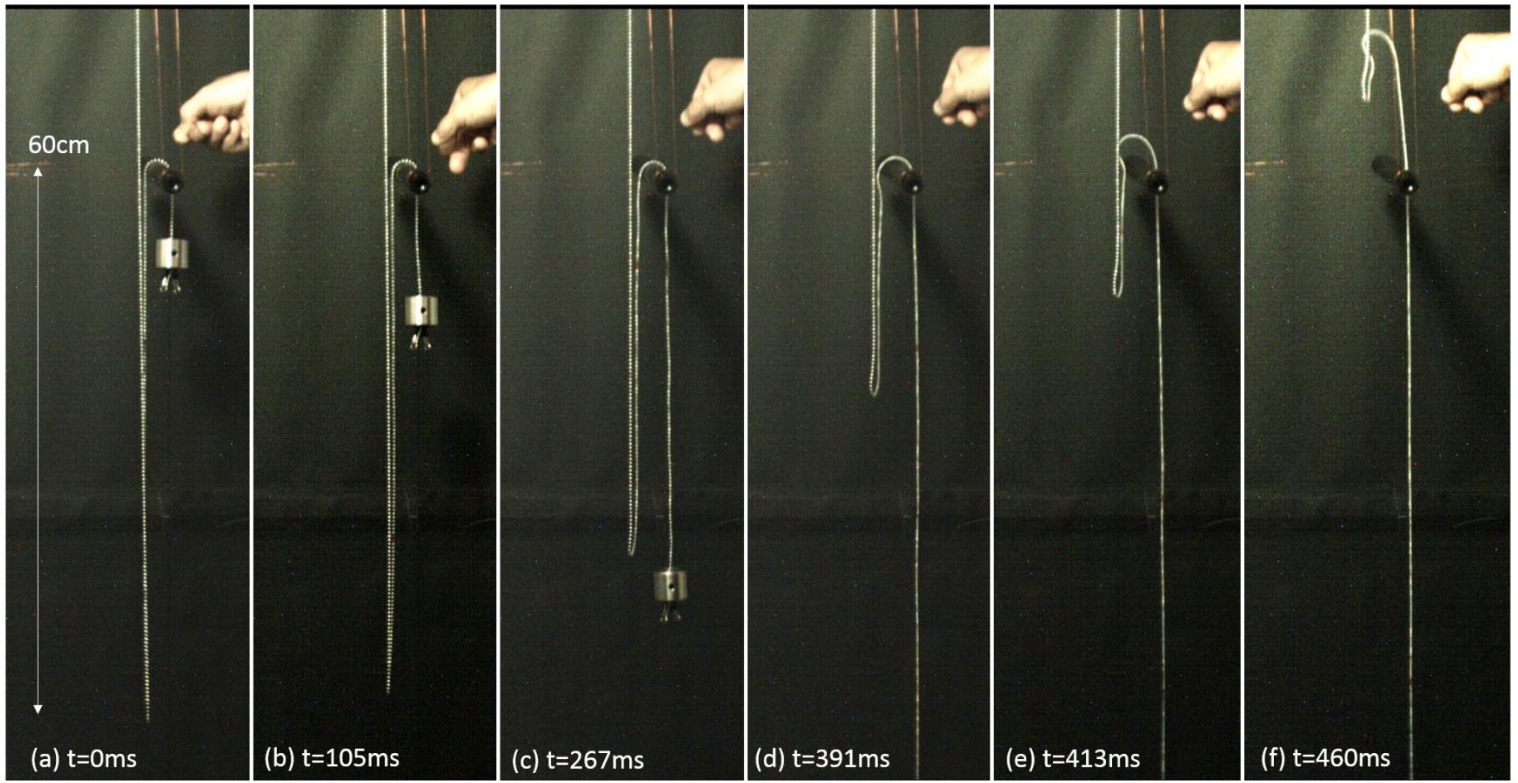

Figure 5. Time laps snapshots of the minimal chain fountain with a 185-cm long chain fixed at the 185cm-high ceiling. A heavy weight was attached to the end of the chain over the black acetal rod. (a) The weight just released. (b) The weight being accelerated, pulling the chain. The bottom of the hanging chain on the left was moving upward at half the speed of the weight. (c) The velocity was close to the threshold velocity. (d) The chain over the sliding rod was lifted up. (e) The fountain grew further while consuming the hanging chain. (f) The fountain rose up at an increasing velocity and disappeared when the weight reached the floor. The threshold velocity was reached at $t$ =360ms ~ 380ms.

https://www.dropbox.com/s/mxxsxn3he7zxww9/Fig%205%20Minimal%20Chain%20Fountain.wmv?dl=0

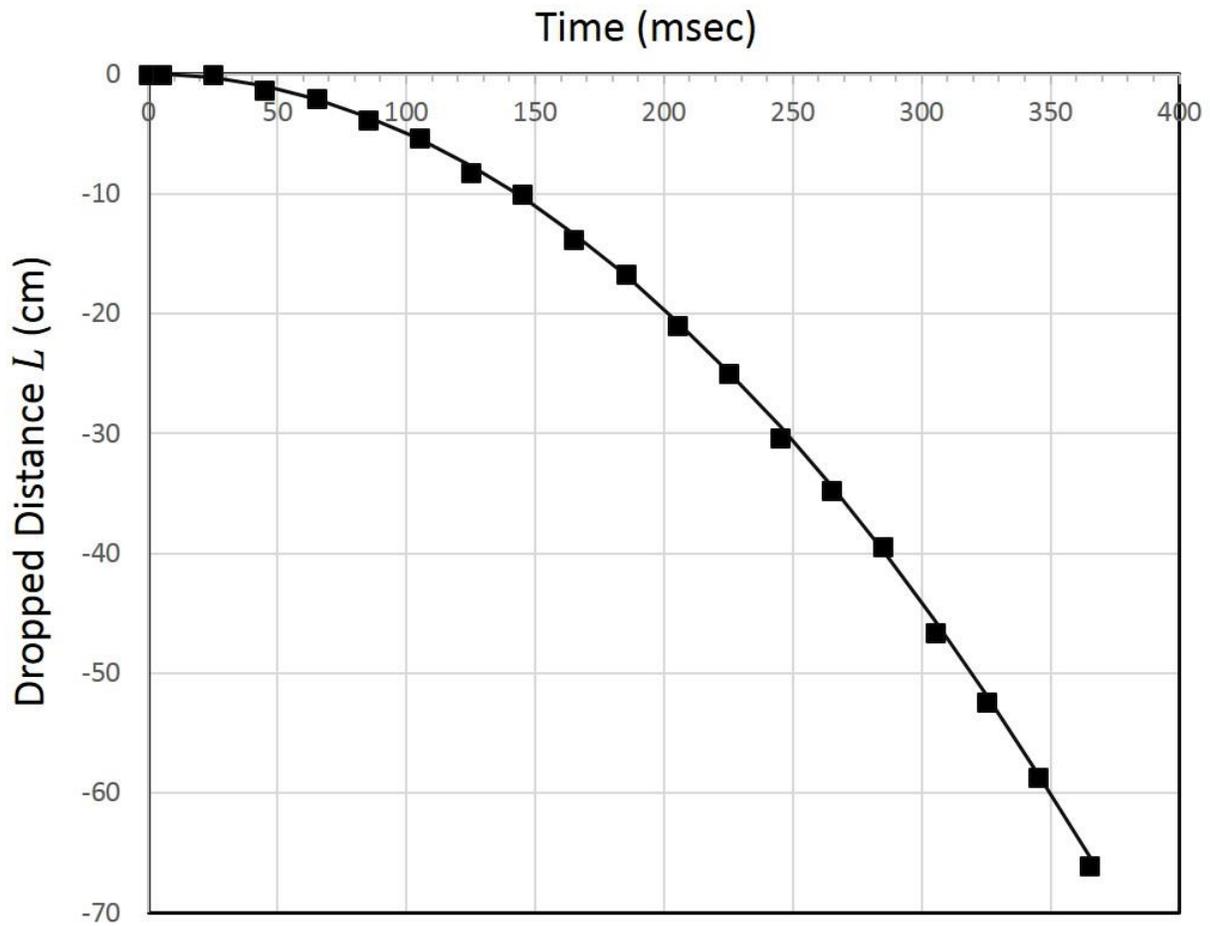

Figure 6.  The dropped distance of the weight as a function of time after release.  The solid line is a theoretical curve for a free fall following $gt^2/2$ with $g$ =9.8m/s² being the gravitational constant.  The solid squares the distance measured from the snapshots taken every 1ms.  Since the mass of the weight is much larger than the chain used, the falling velocity is practically identical to that of free fall.

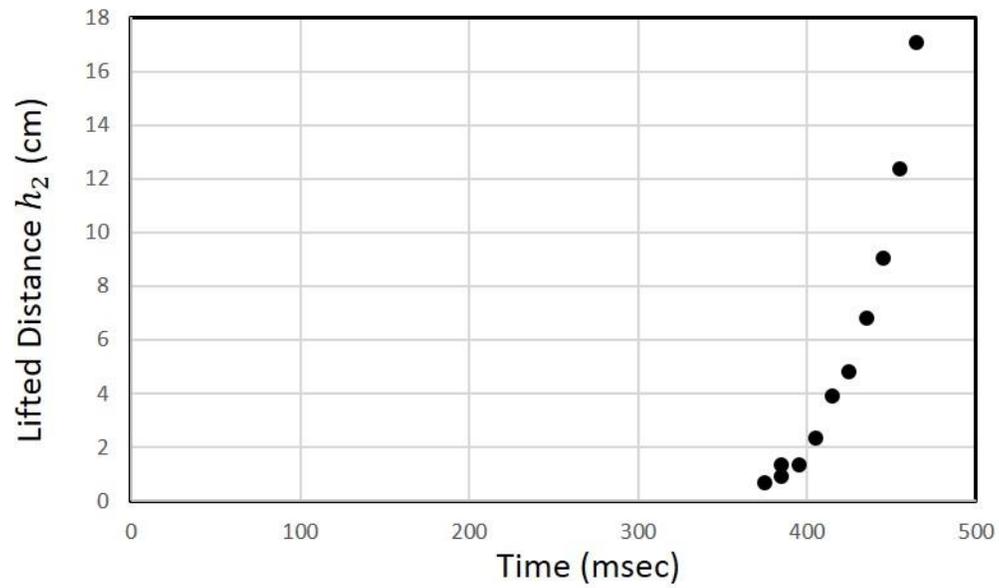

Figure 7. Height of the fountain in the minimal chain fountain as a function of time after release. It clearly indicates the departure of the chain from the sliding rod took place at $t =$360ms $\sim$ 380ms in good agreement with the prediction based on energy conservation.

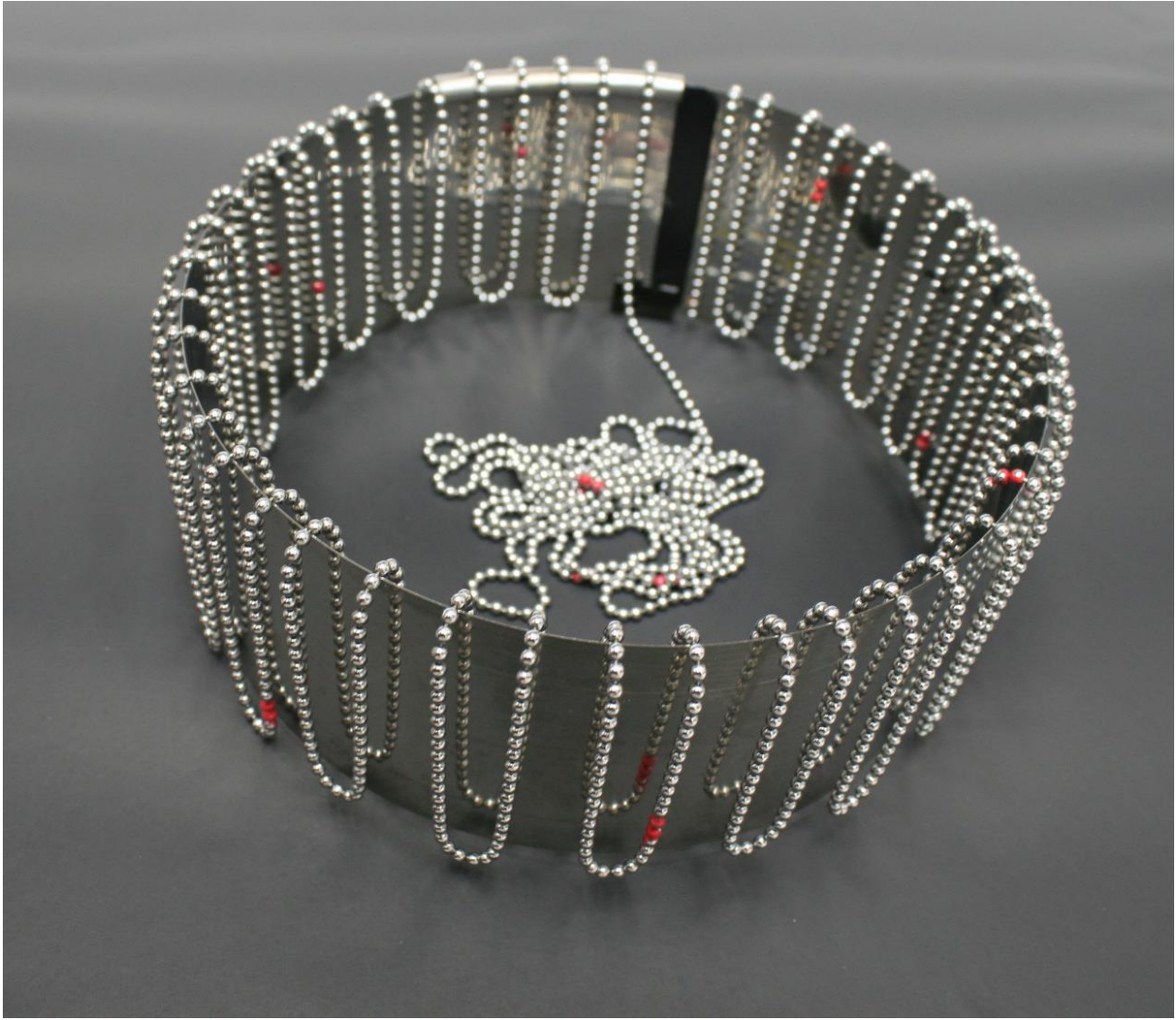

Figure 8. Cascade of hanging chains to achieve a sustained chain fountain without the alleged impulsive kicks from support. An 8cm-wide and 0.25mm-thick stainless steel sheet was used to form the crown whose diameter is 19cm. The chain was supported by the crown only at the linkages between balls by the thin metal sheet, thereby removing a chance for the chain to be kicked upward by the support in the way conjectured by Biggins and Waner[10]. To facilitate a smooth startup of the chain fountain, a 7cm-long and 7mm-diameter stainless steel rod is attached to the rim of the crown to support the four initial hanging chains connected to the rest of the chain to be dropped. At every 50cm, balls were painted red to trace the motion of the chain.

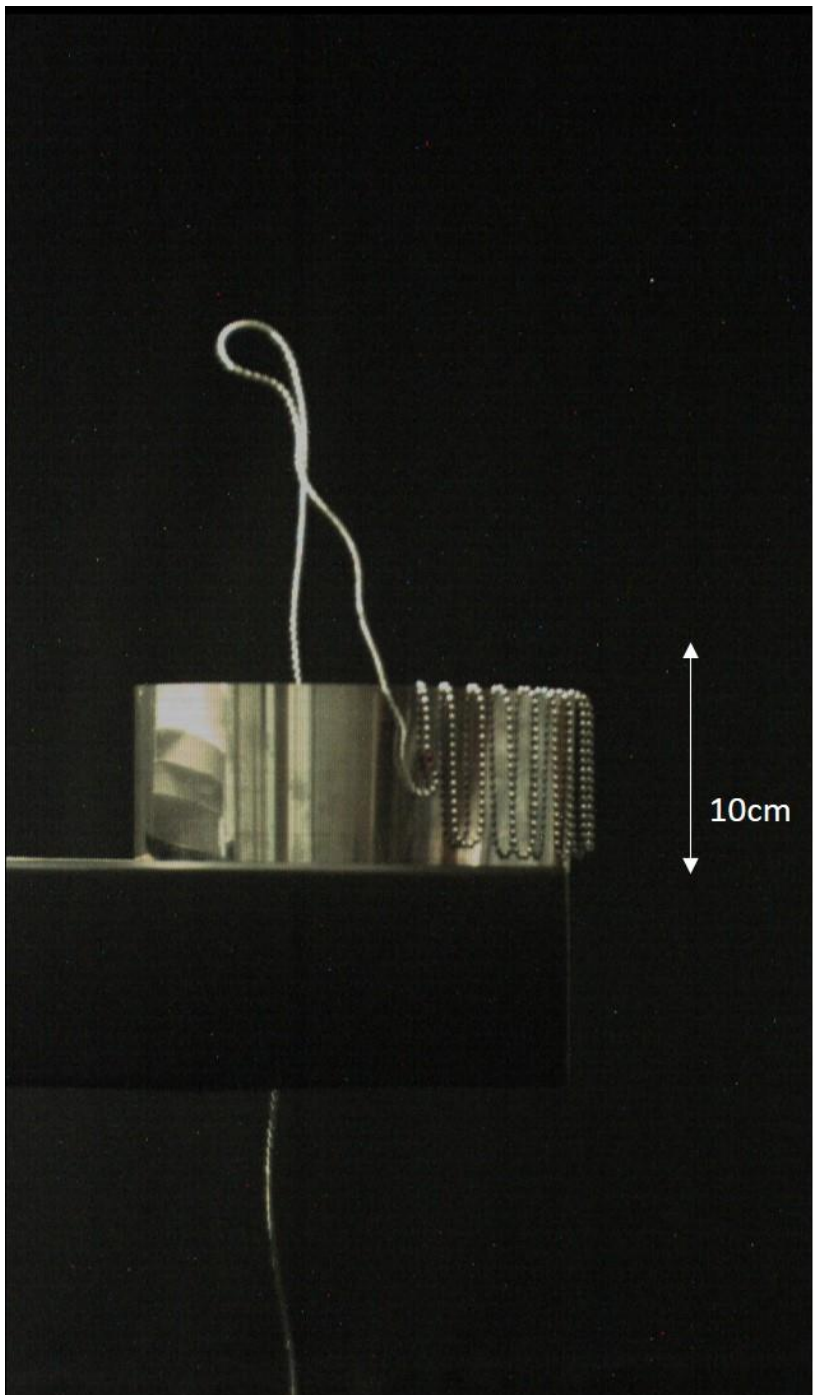

Figure 9. The chain fountain from the crown, which was placed on the same platform as Mould's chain fountain shown in Fig.2. The height fountain consistently exceeded 20cm, a considerably larger value than that for Mould's fountain under the same condition. The complete slow motion video is available as supplemental information.

https://www.dropbox.com/s/7xiyj5jpnzooocd/Fig%209%20Crown%20Fountain.wmv?dl=0

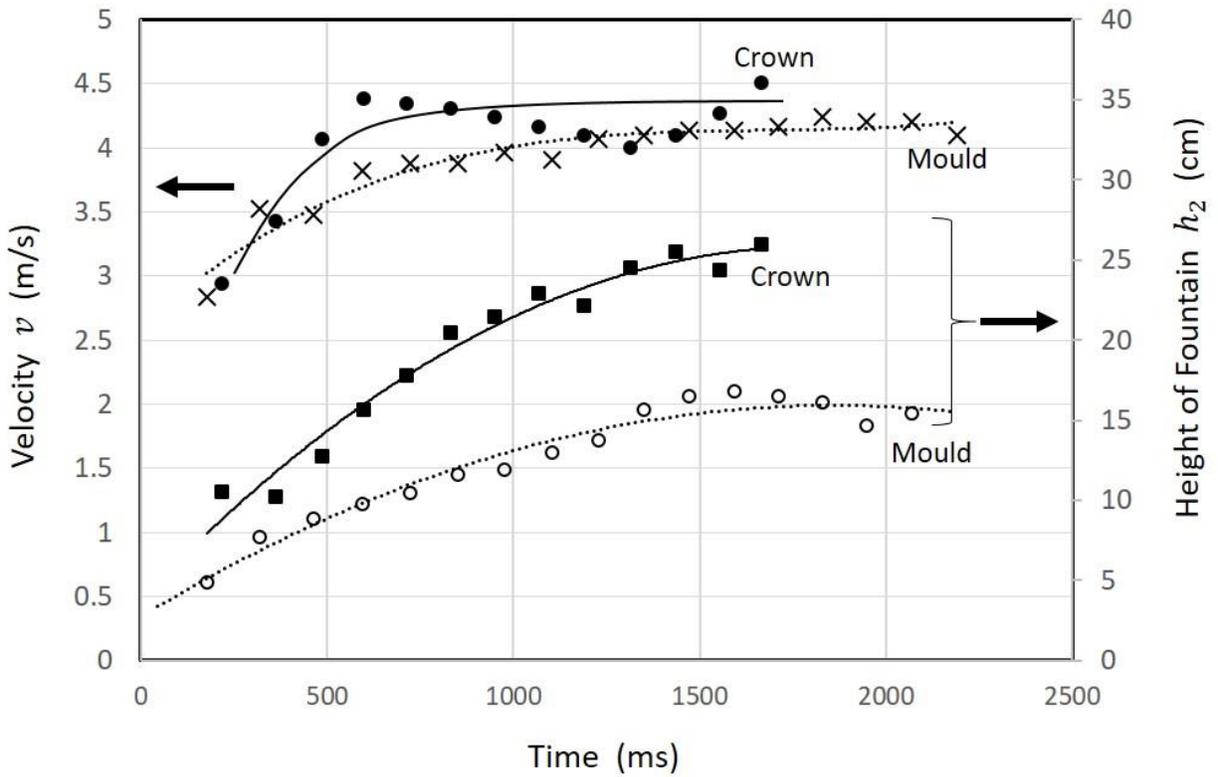

Figure 10. Height of chain fountain and the falling velocity as a function of time. The origin of time was taken as the point where a finite fountain was first observed. There is a significant difference of fountain height for the fountain from the crown support (solid square) and Mould's fountain (open circle), even though the falling velocity of chain was saturated at nearly the same level.

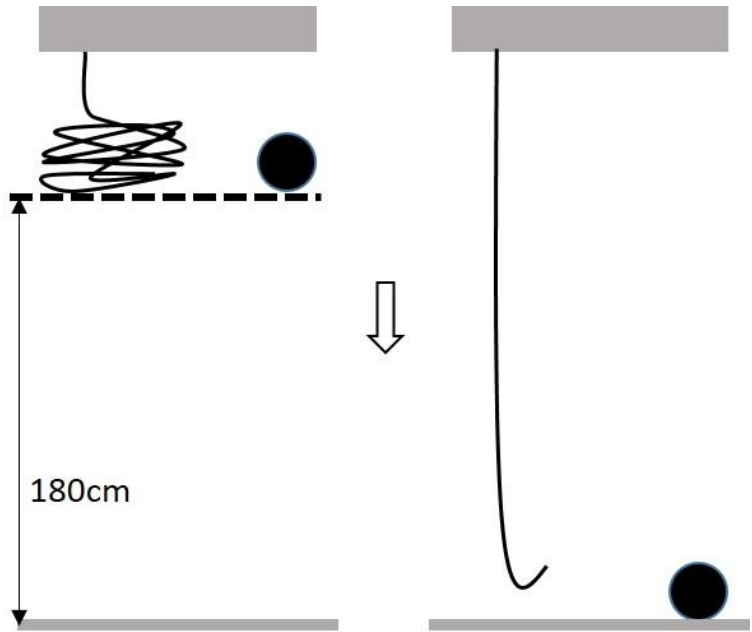

Figure 11. Schematic illustration of free drop experiment of the chain slack. A slack of chain is prepared in the same manner as the case of Mould's chain fountain on a support. A reference aluminum weight is also placed on the support, and the support is quickly pulled down by hand. As the chain falls, the hanging vertical chain is extracted until the end reaches the floor. Observation of the arrival to the floor in reference to the freely falling weight and the motion of the chain after the arrival gives insight about the energy dissipation associated with the extraction of chain from the slack in the absence of alleged upward kicks from the support.

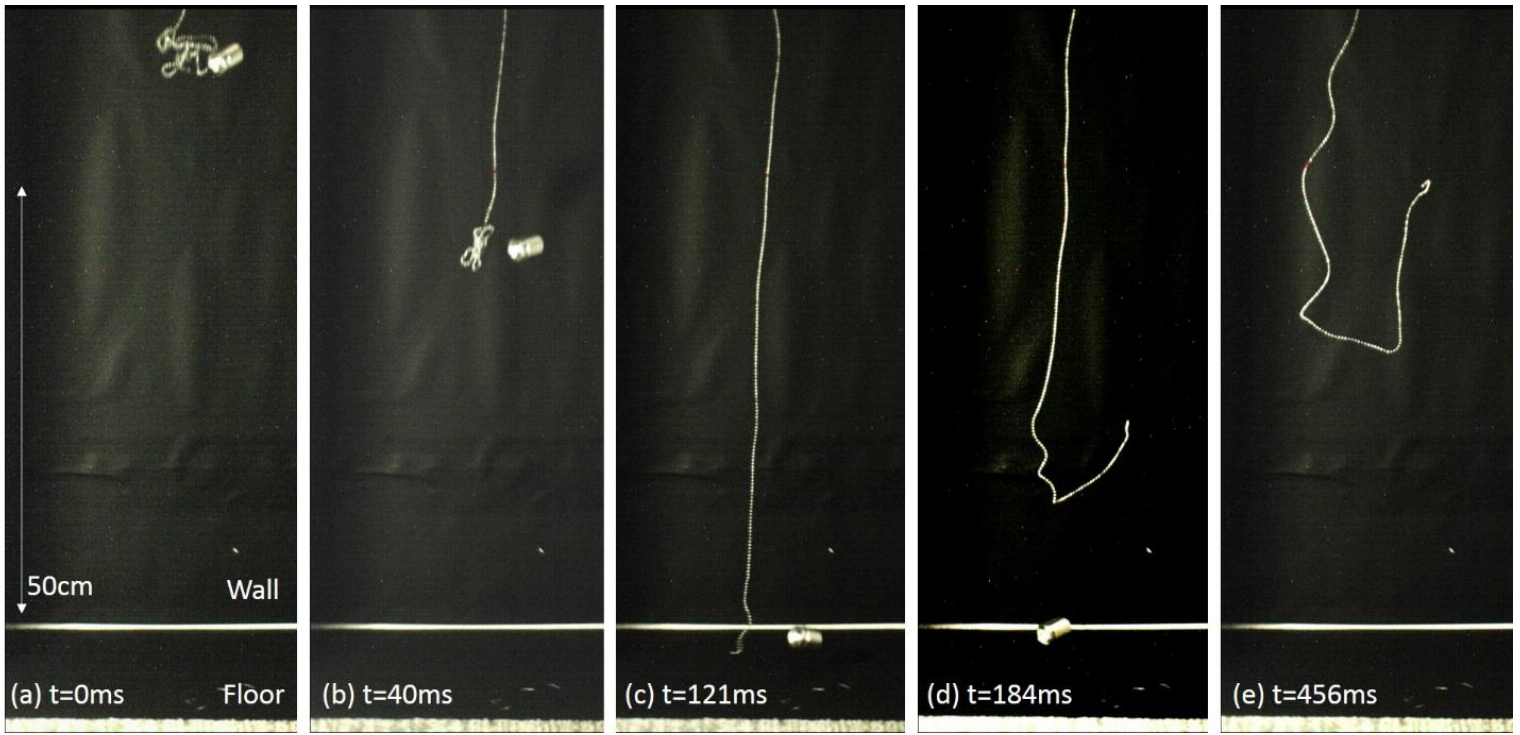

Figure 12. Time laps snapshots of falling chain slack along with a freely falling aluminum weight. The upper end of the 180cm-long chain was fixed at 180cm-high ceiling. The chain was folded into an unentangled slack, which was released to fall freely. The pictures show only the lower part of the fall station from the floor for the limited field of view of the camera. (a) The slack is falling side-by-side with the weight. (b) The fall was continued at the free fall velocity; note that the vertical portion of the chain above the slack is not straight. (c) The fall was complete and the slack was fully extended. The end of the chain reached the floor nearly at the same time as the aluminum weight did. (d) The chain bounced back high. (e) The highest position of the rebounded chain, from which the chain again fell down; Note that the upper part of the chain was also wiggled, indicating the remaining energy was transmitted back high up the chain. The complete slow motion video is available as supplemental information.

https://www.dropbox.com/s/516nvao514tmhrj/Fig%2012%20Falling%20Slack.wmv?dl=0

https://www.dropbox.com/s/din0bx9l70iyp7b/Fig%2012%20Falling%20Slack%2010ms-1sec.wmv?dl=0

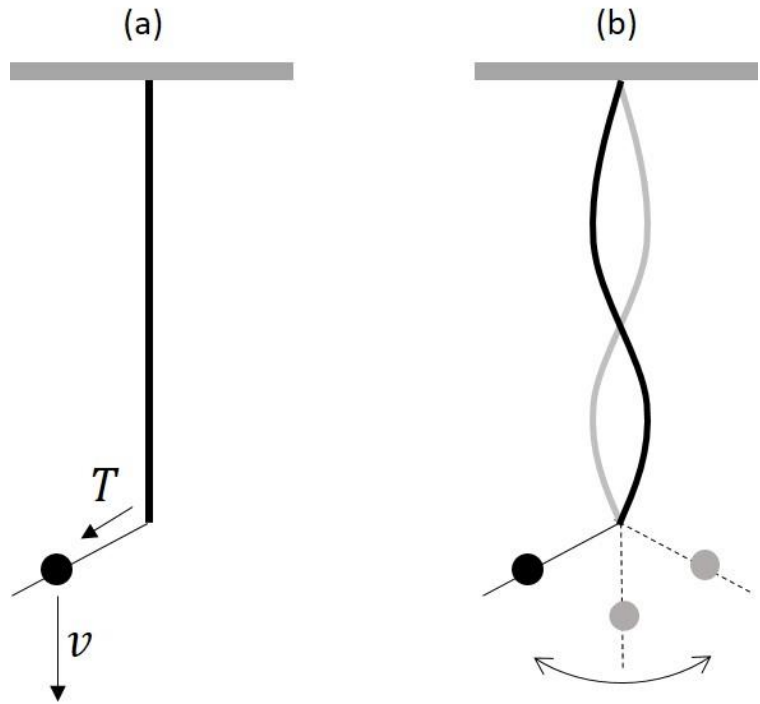

Figure 13. Lateral motion of the hanging chain excited by the laterally extended falling slack. (a) A ball in the chain that is falling not collinearly with the hanging chain exerts a lateral force to the hanging chain, and (b) drives a lateral motion of the hanging part; when the ball is transferred to the hanging part, the kinetic energy along the vertical axis is converted to the kinetic energy in the horizontal direction.

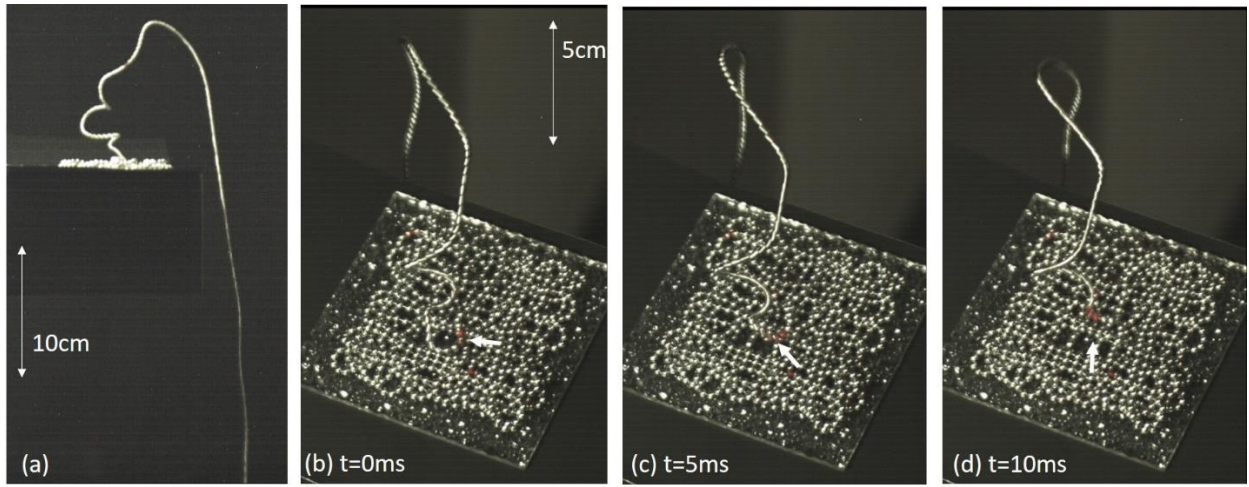

Figure 14. Chain fountain from a tailored slack of chain on a corrugated plastic plate without rim. The chain was arranged in alternating layers of x and y oriented regular lines of chain. (a) Side view of the chain fountain. The uplift point migrated following the chain direction, but was confined in a central region of the chain, presumably due to the partial hindrance of lateral motion of chain. (b)-(d) Snapshots of the slack to indicate the continuous acceleration of balls on the slack before liftoff. The velocity of falling chain is 4.1m/s. Prior to departure, balls were accelerated to at least 1m/s, which is enough to account for the required reduction of energy dissipation. The complete slow motion video is available as supplemental information.

https://www.dropbox.com/s/xu7odyidv2f518n/Fig%2014%20Chain%20Fountain%20from%20XY%20Stack.wmv?dl=0

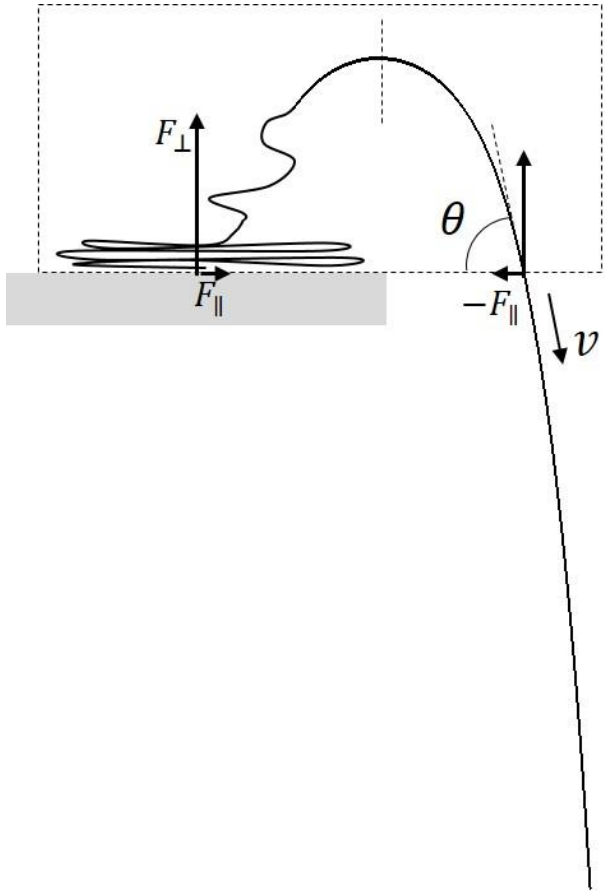

Figure 15. Balance of forces in the chain fountain. The falling side of the fountain is well approximated by the inverted catenary. The upward force from the table is independent of the microscopic details of uplifting. Therefore, the macroscopic force measurement cannot address any microscopic mechanisms.

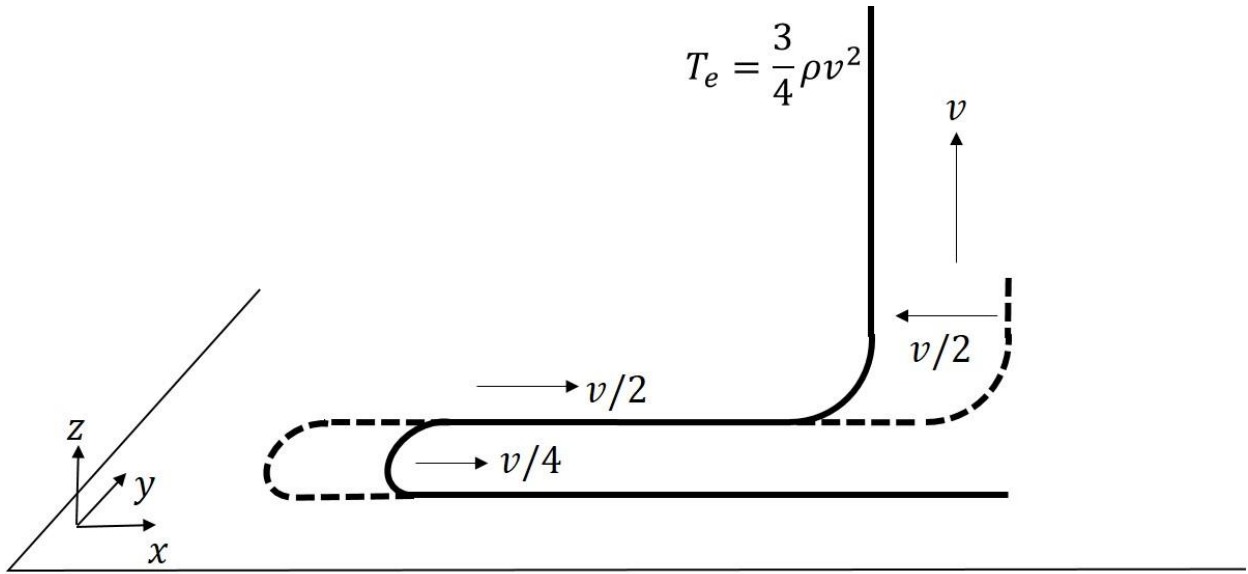

Figure 16. Dissipationless extraction of chain from the slack sitting on a flat table. Based on the concept of the hanging chain, which is an energy conserving system to a good approximation, it is possible to construct a chain extraction scheme without consuming any energy. The chain, sitting still in the front line, is continuously accelerated through a semi-circle first to the velocity $v/2$. The chain moving in plane is picked up vertically at the point moving in the opposite direction at the same velocity. The uplifted chain has velocity $v$ in z-direction, and $-v/2$ in x-direction, so that the total kinetic energy is $3\rho v^2/4$. The assumed precursory acceleration of chain was observed in Fig.14; the motion of the chain shown in Fig.14 is consistent with this scheme.